\documentclass[journal=ancac3,dvipsnames,manuscript=article]{achemso}

\usepackage[T1]{fontenc} % Use modern font encodings
\usepackage{amsmath,amsfonts,amssymb, mathtools}
\usepackage{geometry}
\usepackage{mathrsfs}
\usepackage{physics}
\usepackage{setspace}
\usepackage{bbold}
\usepackage{calc}
\usepackage{subcaption}
\usepackage{array}
\usepackage{textcomp}
\usepackage{gensymb}
\usepackage{subfiles} % include other .tex files in the document
\usepackage{lipsum}  % provides blindtext
\usepackage{siunitx}
\usepackage{booktabs}

\usepackage{color,soul}
\usepackage{xcolor}
\usepackage{comment}
\usepackage{algorithm}
\usepackage{algpseudocode}

\usepackage{xr-hyper}
\usepackage{hyperref}
\usepackage[version=4]{mhchem}

\author{Matteo Cartiglia}
\email{matteopero.cartiglia@imec.be}
\altaffiliation{These authors contributed equally}
\author{Natan Biesmans}
\altaffiliation{These authors contributed equally}
\author{Wannes Peeters}
\author{Wouter Botermans}
\author{Koen Ongena}
\author{Liam Vandekerckhove}
\author{Wouter Renckens}
\author{Eric Beamish}
\affiliation[imec]{imec, Kapeldreef 75, 3001 Heverlee, Belgium}
\author{Elizabeth Skelly}
\author{Kirill A. Afonin}
\affiliation{Chemistry and Nanoscale Science Program, Department of Chemistry, University of North Carolina at Charlotte, Charlotte, North Carolina 28223, United States}
\author{Pol van Dorpe}
\affiliation[imec]{imec, Kapeldreef 75, 3001 Heverlee, Belgium}
\alsoaffiliation{Department of Physics and Astronomy, KU Leuven, Celestijnenlaan 200D, B-3001 Leuven, Belgium}
\author{Sanjin Marion}
\email{sanjin.marion@imec.be}
\affiliation[imec]{imec, Kapeldreef 75, 3001 Heverlee, Belgium}

\title{Data Sieving for Scalable Real-Time Multichannel Nanopore Sensing}

\begin{document}

\begin{abstract}

High-throughput solid-state nanopore experiments generate continuous MHz-rate data streams in which only a small fraction of data contains informative molecular information. This creates storage and processing bottlenecks that limit experimental scalability. 
We introduce Data Sieving, a GPU-accelerated acquisition framework that integrates real-time event detection directly into the measurement pipeline and selectively stores and allows real-time analysis of snapshots around molecular translocations. The system employs a lightweight rolling-average and min–max trigger to identify event candidates in parallel across channels. This architecture reduces stored data volume by up to 98\% while preserving complete molecular signatures across a wide temporal range, from microsecond-scale protein dynamics to second-scale nucleic acid nanoparticle events. 
Continuous baseline monitoring enables autonomous closed-loop actuation; in high-concentration DNA experiments, automatic declogging restored pore conductance, reducing the time spent in a non-productive clogged state to near-zero and without interrupting parallel measurements. 
Validated across DNA, protein, and nucleic acid nanoparticle measurements, Data Sieving links data storage directly to molecular information content rather than experiment duration, enabling scalable, real-time operation of parallel nanopore sensors. The approach provides a hardware-agnostic foundation for long-duration, high-bandwidth single-molecule experiments and other event-driven sensing platforms. 
By using algorithms intrinsically compatible with low-latency digital architectures, this framework provides a clear path toward high-bandwidth, highly multiplexed recording across hundreds of individual nanopore channels in both solid-state and biological pores.

\end{abstract}

\section{Introduction}
\label{introduction}

%\paragraph{Nanopore sensing}

Solid-state nanopores are nanometer-scale holes in thin dielectric membranes that detect individual molecules by monitoring transient blockades of the ionic current~\cite{li2003dna, storm2005fast}. Charged analytes such as DNA, RNA, proteins, or nucleic acid nanoparticles are driven through the pore by an applied electric field and forced to translocate, producing current pulses whose duration and amplitude report on molecular size, structure, and conformation. Advances in chip fabrication, surface chemistry, and electronics over the past decade have substantially increased sensitivity and throughput, expanding nanopore sensing from simple sizing toward probing complex structural transitions and heterogeneous mixtures~\cite{Xue2020, soni2025full, yusko2017real}. A key advantage of nanopores is their ability to detect single molecules in real time without labeling or chemical modification~\cite{Galenkamp2025}, enabling observation of nanoscale processes such as protein folding dynamics and receptor–ligand or antibody–ligand interactions on microsecond timescales~\cite{Zhang2025}.

Nanopore measurements span a broad temporal range depending on molecular dwell time within the sensing region. Motor enzymes can slow molecular motion to millisecond durations that are adequately sampled at kHz bandwidths, enabling commercial DNA sequencing platforms based on large arrays of biological nanopores~\cite{laszlo2014decoding, cherf2012automated, manrao2012reading, clarke2009continuous, brown2016nanopore}. In contrast, freely translocating molecules translocate on microsecond timescales, requiring bandwidths above 1~MHz to resolve transient signals~\cite{li2003dna, storm2005fast, Larkin2014}. This high-bandwidth regime is particularly relevant for proteins, where resolving free translocations or intramolecular dynamics such as side-chain rotations or loop motions demands microsecond temporal resolution~\cite{Larkin2014, Si2017, Schmid2021, Schmid2021b}. In practice, achievable temporal resolution is limited by the interplay between device bandwidth and noise coupling within the nanopore system~\cite{Fragasso2020}.

%\paragraph{Arrays and scalability challenges}

In biosensing, many biomarkers are present at low concentrations, requiring dense nanopore sensor arrays capable of sampling large numbers of molecules and providing immediate analysis to ensure fast time to result~\cite{Ratinho2025}. Achieving adequate sampling therefore depends on recording from large numbers of individually addressable nanopores in parallel. Although substantial progress has been made in developing fluidically and galvanically isolated arrays of solid-state nanopores using microfluidic approaches~\cite{Tahvildari2015, Zhang2015, Jain2017, Tahvildari2017, Rahman2021, Jones2025}, data acquisition and analysis architectures have not advanced at the same pace.

As the number of parallel pores and acquisition bandwidth increase, the resulting data streams rapidly exceed the capacity of conventional acquisition and storage workflows. 
A four-channel system operating at MHz sampling rates already produces hundreds of gigabytes per hour, exceeding 0.8~TB per 30 minutes, while only a small fraction of data points correspond to informative molecular events~\cite{rosenstein2012integrated, fragasso20191, bell2016digitally}. Current experiments typically rely on indiscriminate recording, where all data are written to disk and event detection is performed offline using threshold-based routines, wavelet transforms, cumulative-sum statistics, or hidden Markov models~\cite{shekar2019wavelet, nakane2003nanopore, han2018accurate}. This architecture is fundamentally unscalable: the bottleneck is not analysis sophistication but the requirement that every raw sample be acquired, stored, and transferred before any information can be extracted. Triggering strategies and real-time voltage actuation have addressed specific aspects of this problem~\cite{Langecker2015, Rand2022, Mi2024}, and adaptive sampling has been demonstrated for biological nanopores~\cite{kovaka2021targeted, ulrich2022readbouncer}, but no existing framework provides a scalable, hardware-agnostic architecture for parallel event detection and selective data retention across solid-state nanopore arrays operating at MHz bandwidths.

In this work, we present ``Data Sieving,'' a scalable acquisition framework applicable to both biological and solid-state nanopores that resolves this architectural bottleneck. The central design constraint is scalability to multichannel, MHz-sampling-rate experiments: triggering algorithms must be computationally lightweight and ultra-parallelizable to support real-time, high-throughput data acquisition.

Data Sieving performs early event detection at the edge of the measurement system, selectively storing only short, full-resolution snapshots around molecular translocation events. 
This (a) links stored data volume directly to molecular information content rather than experiment duration; (b) accelerates analysis by performing it in real-time near the point of acquisition; and (c) passes forward in the pipeline only useful parts of the current traces for further analysis. 
The result is a scalable, edge-computing architecture that operates within the data transfer limits of modern computing interfaces.

\begin{figure*}
    \centering
    \includegraphics[width=\linewidth]{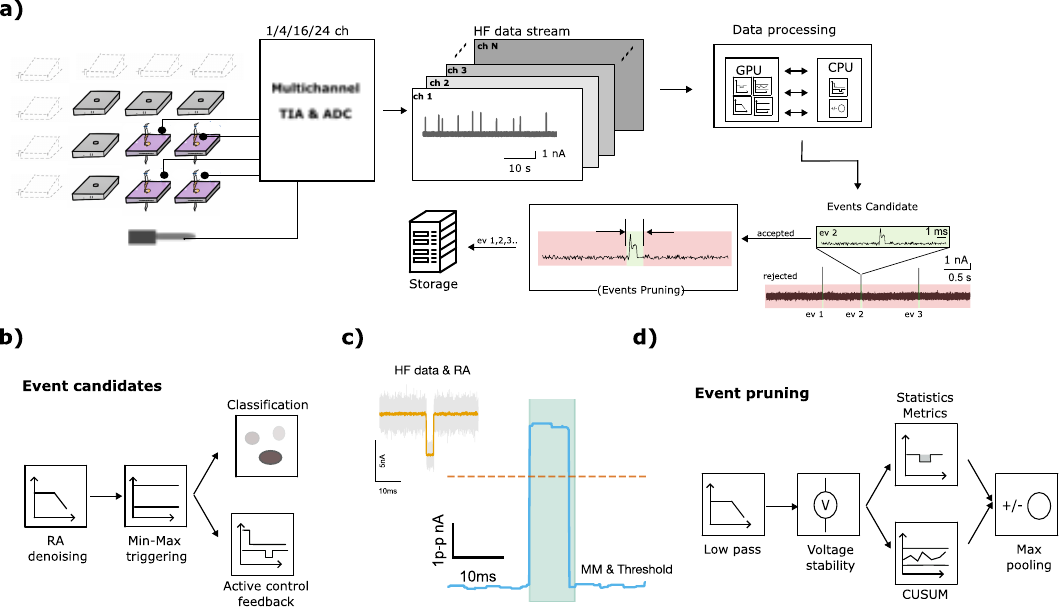}
\caption[]{\textbf{Data Sieving architecture and algorithmic pipeline.} 
\textbf{(a)} System Overview: High-frequency (HF) data from nanopores is recorded via multichannel amplifiers (1, 4, 16, or 24 channels). Data is processed using a heterogeneous CPU/GPU strategy to manage real-time throughput. The end result of the pipeline is the extraction of event candidates for further processing of the parallel high-bandwidth data streams. These event candidates are then further "pruned" to remove any unnecessary padding around them, with the intent of reducing data storage and real-time processing requirements later in the pipeline. 
\textbf{(b)} Event Candidate Detection: Parallelized detection uses a rolling-average (RA) filter for denoising and windowed min-max (MM) triggering. This stage simultaneously enables active-feedback control and coarse signal classification. Thresholds are automatically initialized from baseline noise statistics, though they can be manually overridden.
\textbf{(c)} RA+MM Logic: The detector acts as a tunable band-pass filter where the RA suppresses high-frequency noise and the MM rejects slow baseline drift by evaluating local peak-to-peak amplitude. 
\textbf{(d)} Event Pruning: Raw candidate snapshots undergo digital low-pass filtering and voltage-stability checks. Precise event boundaries are localized via parallel metric extraction and cumulative-sum algorithms. If both methods reach a consensus, the event is trimmed to the agreed interval; otherwise, the full untrimmed snapshot is retained to prevent information loss. }
\label{fig:fig1}
\end{figure*}

The Data Sieving framework, illustrated in Fig.\ \ref{fig:fig1}a, adopts an "edge-computing" paradigm by integrating real-time detection and automated actuation directly into the acquisition pipeline. The core of this system is a parallelized event-candidate detection stage (Fig.\ \ref{fig:fig1}b) that uses a rolling-average (RA) and windowed min–max (MM) trigger. This architecture functions as a tunable band-pass filter (Fig.\ \ref{fig:fig1}c) that rejects high-frequency noise and slow baseline drift, enabling high-sensitivity capture while remaining computationally lightweight for GPU execution across many channels. Full algorithmic details and parameter selections are provided in the Methods section and in Supplementary Material Sections~\ref{sm:ec} and~\ref{sm:thoretical_bandpass}.

Beyond detection, the system utilizes the downsampled baseline to maintain a lightweight health monitor for each channel, enabling autonomous closed-loop actuation—such as rapid polarity inversions for automatic declogging—without interrupting parallel measurements. To further enhance acquisition efficiency, an automated event pruning stage (Fig.\ \ref{fig:fig1}d) utilizes two independent event detection methods to trim unnecessary baseline padding. 
By combining these real-time filtering and recovery mechanisms, Data Sieving reduces stored data volume by up to 98\% while preserving complete molecular signatures across a wide temporal range. In the following sections, we quantify the performance and scalability of this framework across diverse molecular assays.

\section{Results}

\subsection{Experimental Validation and Data Sieving Performance}

To validate the real-time detection capabilities of Data Sieving, we conducted simultaneous four-channel experiments using solid-state nanopores (9–12.1~nm diameter) with a voltage of 400~mV in 4~M LiCl (17~S/m). We benchmarked the system using double-stranded (ds)DNA of varying lengths (250 base pairs (bp), 2.5~kbp, and 10~kbp) to span a wide dynamic range of translocation kinetics. Event candidates were detected in real time using the RA+MM trigger architecture, with both raw snapshots and pruned events stored for downstream analysis.

\begin{figure*}
\centering
\includegraphics[width=1\linewidth]{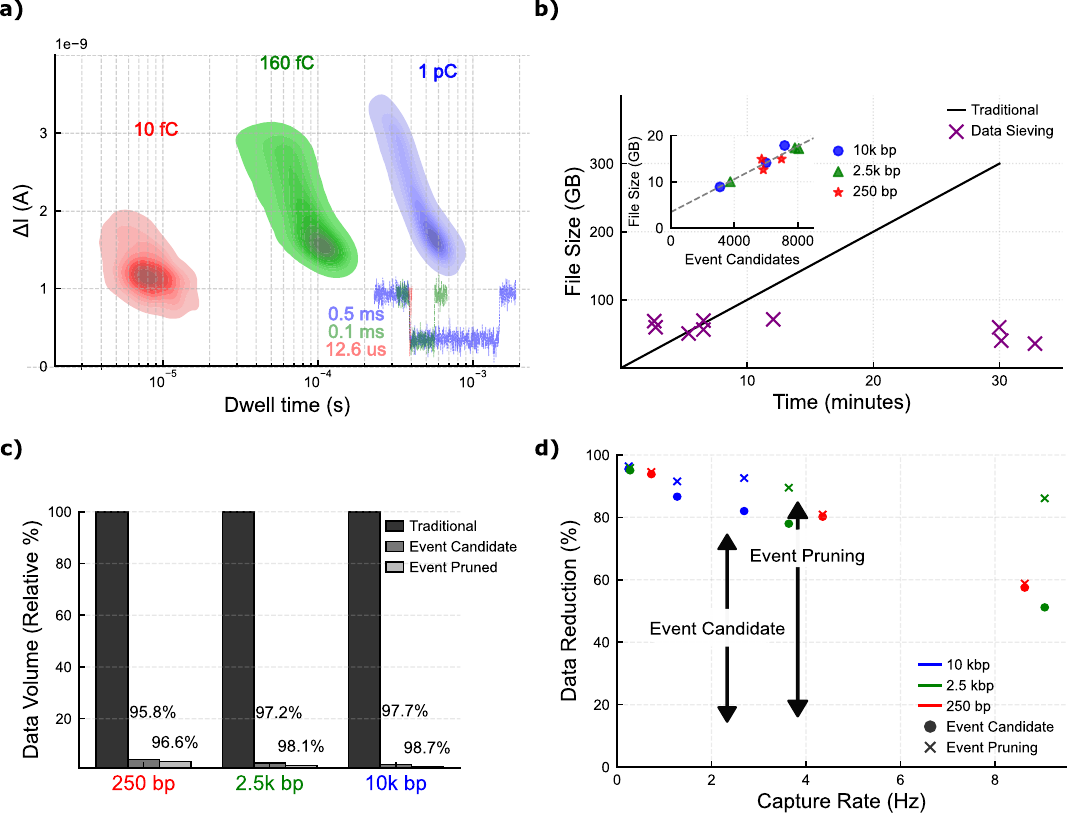}
\caption{\textbf{Experimental validation and data reduction efficiency.} All measurements at 400~mV in 4~M LiCl (17~S/m), four channels at 27~MHz, pore diameters 9--12.1~nm.
    \textbf{(a)} Density plots of mean current drop ($\Delta I$) versus dwell time for translocation events corresponding with the passage of 250 bp (red), 2.5k bp (green), and 10k bp (blue) dsDNA. Integrated charge values and representative traces confirm high-fidelity capture across three orders of magnitude in dwell time.
    \textbf{(b)} Comparison of storage growth: traditional logging (solid line) versus Data Sieving (purple markers) for 250~bp, 2.5~kbp, and 10~kbp dsDNA at three concentrations, run until $\sim$1000 events per channel were detected across four channels. Data Sieving links file size to molecular information content rather than experiment duration. The inset shows a linear relationship between file size and the number of detected events, regardless of experimental duration. 
    \textbf{(c)} Relative data volume reduction for each molecular species. Massive reduction from traditional recording to event candidate storage ($\sim$95\%) and final pruning ($>$98\%). 
     \textbf{(d)} Reduction efficiency as a function of capture rate. Data Sieving maintains $>$80\% reduction at high event frequencies.}
\label{fig:fig3_datareduction}
\end{figure*}

Figure~\ref{fig:fig3_datareduction}a demonstrates that the framework reliably captures molecular signatures across three orders of magnitude in dwell time. The density plots of mean current drop ($\Delta I$) versus dwell time show distinct clusters for each DNA length. The integrated charge values range from $\sim$10~fC for 250~bp up to $\sim$1~pC for 10~kbp dsDNA. Representative traces, aligned to event onset, show that the system faithfully preserves the structural information of both fast ($\sim$12.6~$\mu$s) and long-duration ($\sim$0.5~ms) translocations. 

The primary advantage of the data sieving approach is a massive reduction in stored data volume. In a standard 30-minute acquisition at 4~$\times$~27~MHz (16-bit), traditional indiscriminate logging would produce $\sim$0.8~TB of data including voltage traces. By contrast, Data Sieving links file growth directly to molecular information content rather than experimental duration. As shown in Fig.~\ref{fig:fig3_datareduction}b, while traditional logging grows linearly over time, Data Sieving storage scales only with the number of detected event candidates.

The magnitude of this reduction is quantified across species in Fig.~\ref{fig:fig3_datareduction}c. Logarithmic scaling highlights that storing only event candidates reduces raw data volume by $\sim$95\%. The application of event pruning further enhances efficiency, removing redundant baseline padding to bring the data reductions exceeding 98\% for 10~kbp dsDNA. 

Finally, we evaluated the scalability of this reduction under varying experimental loads. Figure~\ref{fig:fig3_datareduction}d shows that even as the capture rate increases, the framework consistently maintains a high reduction efficiency. In this demonstration, the event pruning algorithm utilizes a 50~kHz low-pass filter; while this setting is well-tuned for the sustained signals of larger molecules ($>$2.5~kbp), it was left unchanged specifically to showcase its performance across different molecules. Because 250~bp dsDNA translocates so rapidly ($\sim$12.6~$\mu$s), this filter cancels the fast signal transitions, which explains the lower pruning effectiveness for 250~bp dsDNA (in red). 

The RA and MM windows define the effective band-pass of the detector: the RA window sets the high-frequency cutoff and must be tuned to the expected translocation timescale, while the MM window sets the low-frequency cutoff and temporal length of stored snapshots, balanced against detection latency. Crucially, the RA+MM detector is highly tolerant:  robust detection is achieved across a broad range of parameter settings without fine-tuning, facilitating deployment across diverse experimental conditions. Detailed theoretical derivations are provided in Supplementary Material Section~\ref{sm:thoretical_bandpass}.

\subsection{Active Feedback and Fast Actuation}

Clogging events are a primary hurdle to nanopore performance. They increase baseline noise, cause unstable signals, and will prematurely terminate an experiment if not resolved quickly. 
Data Sieving leverages its continuous MM monitoring to integrate a fast, autonomous actuation mechanism that responds in real time to sustained deviations.

\begin{figure*}
\centering
\includegraphics[width=1\linewidth]{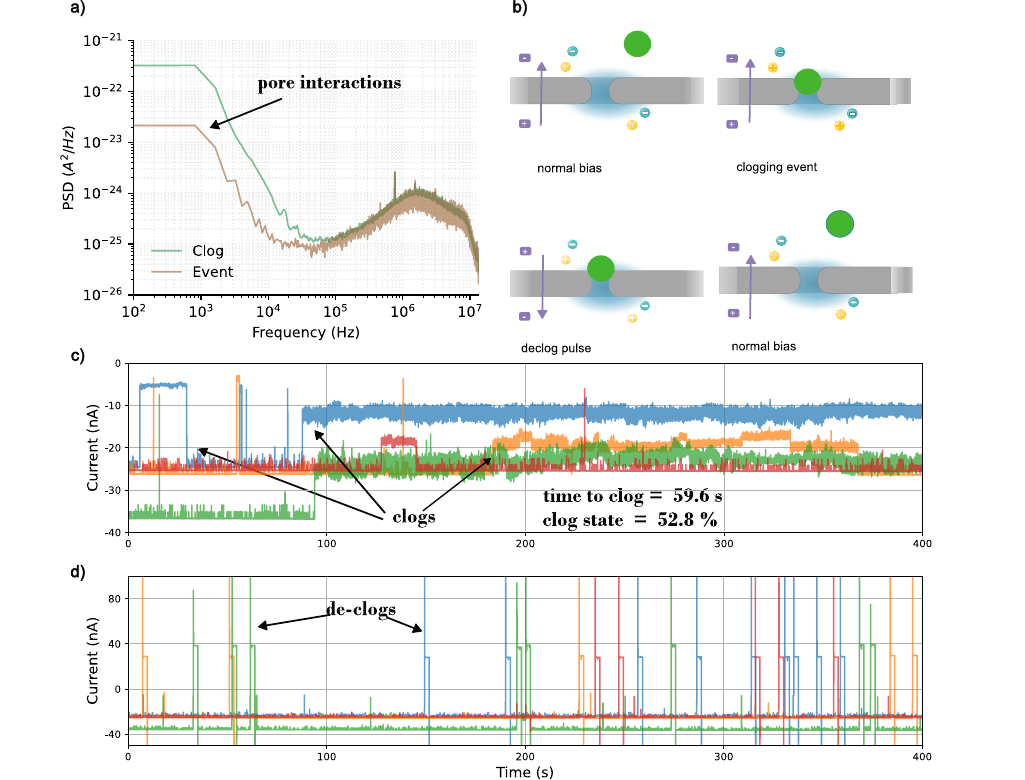}
\caption{\textbf{Real-time pore health monitoring and automatic declogging.} Experiments with 10~kbp dsDNA at 1.8~nM, 400~mV in 4~M LiCl.
    \textbf{(a)} Power Spectral Density (PSD) comparing a normal translocation event to a clogging event. Clogging events are characterized by distinct low-frequency noise, allowing for real-time discrimination.
    \textbf{(b)} Schematic of the automatic declogging cycle: detection of a clog triggers a brief polarity inversion (+600 mV, 1 s) to restore open-pore conductance. 
    \textbf{(c)} Experimental data without active feedback (10 kbp dsDNA, 1.8 nM): Pores are quickly rendered unusable due to frequent clogging, resulting in a 52.8\% clogged state and a mean time to first clog of 59.6 s. 
    \textbf{(d)} Experimental data with automatic declogging enabled (10 kbp dsDNA, 1.8 nM): Data Sieving identifies baseline deviations in real time and issues brief polarity-inversion pulses (+600 mV, 1 s) to restore open-pore conductance. This produces repeated, rapid de-clog events that extend experimental time and data.}
\label{fig:fig_declog}
\end{figure*}

The spectral signatures of clogging and translocation events are compared in Fig.~\ref{fig:fig_declog}a. Power Spectral Density (PSD) analysis shows that clogging produces a distinct noise profile, which differs significantly from a translocation event. These unique spectral fingerprints allow the RA+MM algorithm to accurately discriminate between true translocation signals and pore blockages in real time.

The operational protocol for automatic declogging is illustrated in Fig.~\ref{fig:fig_declog}b. Upon detecting a clog, the system triggers a brief polarity inversion to dislodge the contaminant from the pore. If the recovery attempt is unsuccessful after multiple attempts, the system automatically deactivates the channel to prevent unnecessary crosstalk between channels.

We benchmarked this active feedback using high-concentration solutions (10 kbp dsDNA at 1.8 nM). In experiments without active feedback (Fig.~\ref{fig:fig_declog}c), pores became unusable almost immediately; the mean time to the first clog was just 59.6 s, with channels spending 52.8\% of the total 400 s runtime in a non-productive clogged state.

In contrast, enabling automatic declogging drastically improved experimental continuity (Fig.~\ref{fig:fig_declog}d). Upon identifying a faulty baseline, Data Sieving issues a channel-specific 1 s polarity inversion (+600 mV) to dislodge trapped aggregates. Crucially, this feedback is asynchronous; declogging pulses are applied only to the affected channel, while adjacent channels continue acquiring molecular events uninterrupted. This closed-loop recovery effectively extends pore lifetime and ensures that acquisition remains productive throughout long-duration experiments.

\subsection{Resource Utilization}

The Data Sieving framework is engineered to be computationally lightweight, facilitating deployment on standard general-purpose GPUs. All performance metrics reported here correspond to the real-time processing of four nanopore channels sampled at 27~MHz each, using an RTX~4080~Super as the reference hardware. 
Computational load splits into two independent streams: GPU utilization scales with total incoming data rate (sampling frequency $\times$ channel count), while memory-controller utilization scales with event capture rate, reflecting the overhead transfer of raw snapshots to GPU memory only when an event is detected.

\begin{figure*}[htbp]
\centering
\includegraphics[width=1\linewidth]{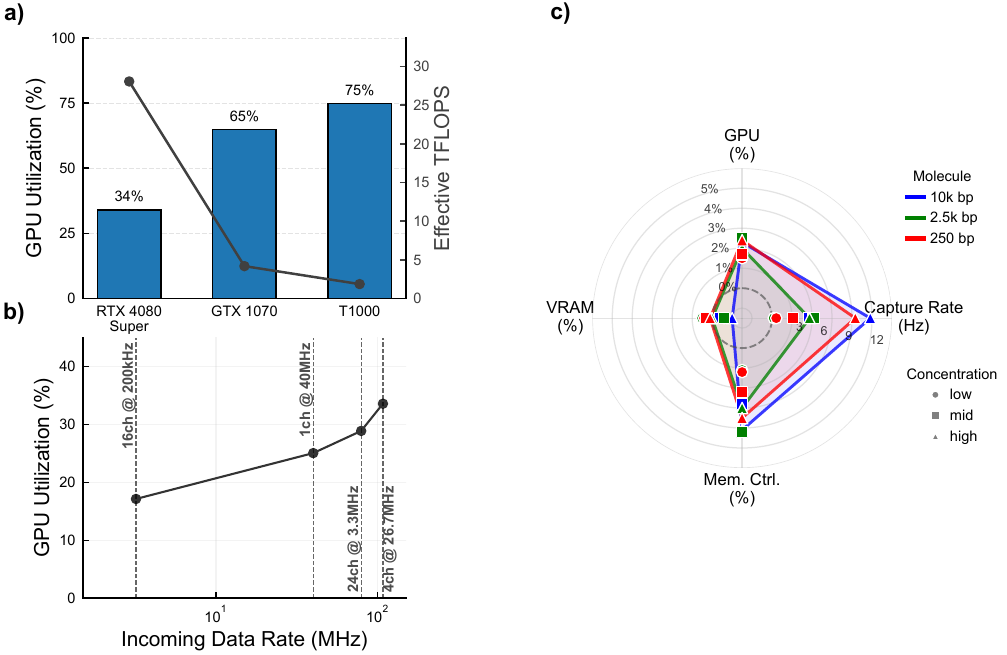}
\caption{\textbf{GPU resource utilization and computational scalability.} Four channels at 27~MHz, 400~mV in 4~M LiCl.
    \textbf{(a)} Hardware benchmarking: Comparative GPU utilization and effective TFLOPS for the parallel acquisition of 4 channels (2.5~kbp DNA at 10~nM) using high-end (RTX 4080 Super), mid-range (GTX 1070), and entry-level (T1000) GPUs. While utilization increases on lower-tier hardware, all models maintain stable real-time acquisition. 
\textbf{(b)} Mean GPU utilization as a function of aggregate incoming open pore data rate. When event-candidate detection is performed directly on the GPU , utilization remains below 40\% for data rates exceeding 100~MHz, indicating computational headroom. 
\textbf{(c)} Radar plot showing incremental GPU utilization, VRAM usage, and memory controller load relative to a control solution without analytes, evaluated across three DNA lengths (250~bp, 2.5~kbp, 10~kbp) and three concentration tiers. Only the memory controller load increases significantly with capture rate, reflecting higher load from raw snapshot transfers, while GPU and VRAM usage remain modest.}
\label{fig:gpu_fig}
\end{figure*}

The impact of active molecular acquisition on system resources is detailed in the radar analysis of Figure~\ref{fig:gpu_fig}c. By comparing experimental runs across various DNA lengths and concentrations to a control baseline, we find that the incremental increase in GPU utilization is remarkably modest. Notably, the primary scaling factor is the memory controller load, which rises in tandem with the translocation capture rate.

Hardware compatibility was further evaluated by benchmarking three GPU generations (Figure~\ref{fig:gpu_fig}b). Both the high-end RTX~4080~Super and the mid-range GTX~1070 handled the four-channel workload with significant headroom (34\% and 65\% utilization, respectively). Even the entry-level T1000, despite its lower core count and memory bandwidth, maintained stable real-time acquisition at 75\% utilization.

Finally, the system's capacity for future high-throughput scaling was tested by measuring GPU load against increasing aggregate data rates (Figure~\ref{fig:gpu_fig}b). Benchmarks spanned from a 16-channel setup at 200~kHz to a four-channel setup at 27~MHz. Extrapolation of this data indicates that the RTX~4080~Super can manage aggregate rates exceeding 100~MHz while remaining below 40\% utilization. This substantial computational headroom ensures that Data Sieving is prepared for next-generation nanopore arrays with significantly higher channel counts.

\subsection{Molecular applications}

To conclude the experimental validation, we demonstrate the versatility of Data Sieving across diverse biomolecular systems characterized by contrasting kinetic profiles. Data Sieving's flexibility is highlighted through two distinct assays: the structural denaturation of the protein streptavidin and the voltage-dependent translocation of nucleic acid nanoparticles (NANPs)~\cite{alibakhshi2017picomolar,pandey2025chemical, panigaj2026expanding,grabow2011self}.

\begin{figure*}[htbp]
  \centering
   \includegraphics[width=\linewidth]{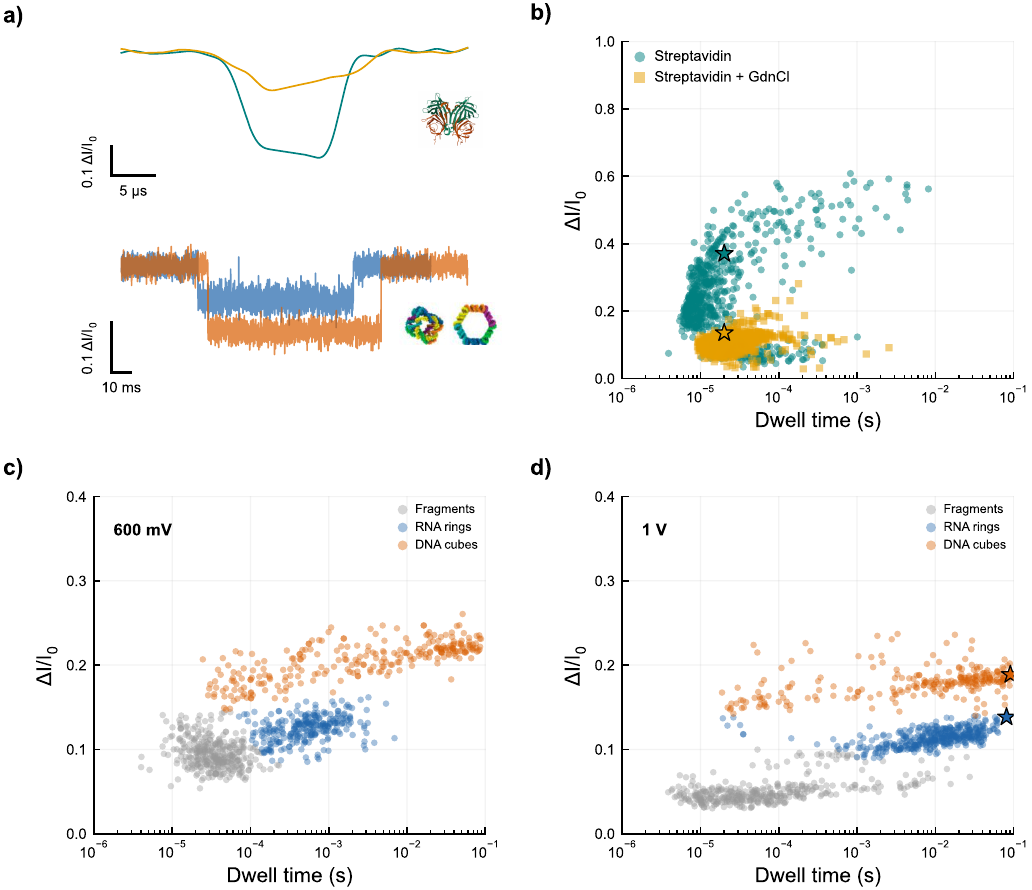}
  \caption{\textbf{High temporal dynamic range biomolecular sensing.}
   \textbf{(a)} Representative current traces spanning the full dynamic range of the system. Top: 120~nM streptavidin under native (2~M KCl, 10~mM Tris) and denaturing (3~M GdmCl, 1~M KCl, 10~mM Tris) conditions, illustrating fast transient translocations in the 10~\textmu s regime. Bottom: nucleic acid nanoparticle (NANP) translocations at 1~V, showing sustained events lasting up to $\sim$100~ms. Scale bars indicate time and amplitude.
    \textbf{(b)} Scatter plot of fractional blockade ($\Delta I/I_0$) versus dwell time for streptavidin. Starred markers indicate the events shown as traces in~(a). The system captures conformational differences between native and unfolded protein states.
    \textbf{(c,\,d)} Mixed nucleic acid nanostructures (1~nM DNA cubes and 0.2~nM RNA rings) at 600~mV~\textbf{(c)} and 1~V~\textbf{(d)} in 0.4~M KCl, 2~mM MgCl$_2$, 10~mM Tris; pore diameter 10.4~nm, 26.7~MHz sampling. Gaussian Mixture Model (GMM) clustering reveals three distinct populations (fragments, RNA rings, and DNA cubes), demonstrating Data Sieving's ability to maintain high-resolution capture over five orders of magnitude in dwell time.}
  \label{fig:fig_applications}
  \end{figure*}

In the first assay (Fig.~\ref{fig:fig_applications}a), we monitored streptavidin under native and denaturing conditions simultaneously across two channels. The resulting distributions of fractional blockade ($\Delta I/I_0$) versus dwell time clearly distinguish the two populations: compact native proteins produce deeper, longer blockades, while unfolded proteins result in shallower, shorter signals. 
 With event durations pushing the fast end of the dynamic range into the  10--100~$\mu$s regime, these results confirm the sensitivity of the RA+MM detector to fast, transient protein translocations, achieved by decreasing the RA window to 10~$\mu$s
 for a high-frequency response.

The second assay (Fig.~\ref{fig:fig_applications}b and c) explores the opposite temporal extreme using representative NANPs, such as six-stranded DNA cubes and hexagonal RNA rings. These NANPs exhibit slow, mechanically hindered translocations within the "mechanical clamp" regime, with dwell times extending from several milliseconds up to $\sim$100~ms. To record these long-duration signals, the MM window was increased to 200~ms to capture the full event duration. A 2-D Gaussian Mixture Model (GMM) clustering reveals distinct signatures for each nanostructure, where rigid DNA cubes produce deep, smooth blockades and flexible RNA rings yield shallower, more variable signals. Additional information is available in Supplementary Information~\ref{sec:sm_nanps_sequence} and \ref{SM:nanoparticles}.

Together, these applications demonstrate that Data Sieving can reliably detect and classify molecular events spanning five orders of magnitude in duration (10~$\mu$s to $\sim$100~ms). This broad dynamic range ensures that the same real-time detection architecture, once the RA and MM parameters are adjusted, is applicable to a wide variety of high-bandwidth sensing platforms, from high-speed protein dynamics to complex nanoparticle characterization.

\section{Conclusions} 

Data Sieving enables stable, real-time reduction and analysis of MHz-rate multichannel nanopore data by integrating event detection directly into the acquisition front-end, filtering more than 98\% of non-informative baseline while preserving complete molecular signatures across analytes spanning five orders of magnitude in dwell time. 
Beyond $\sim$10~MS/s aggregate data rates, classical CPU-based triggering becomes computationally prohibitive, forcing either event loss or full-stream recording; in the latter case, storage bandwidth alone limits experiments to a handful of parallel channels at 20~MS/s each on typical high-performance drives. 
By relocating coarse classification to parallelizable computing units: GPUs today, FPGAs or ASICs tomorrow, Data Sieving sustains real-time performance approaching 100~MHz aggregate rates and eliminates the primary I/O bottleneck that has constrained high-bandwidth nanopore studies. 
The architecture scales hierarchically: by clustering nanopore sensors and applying data sieving across array segments, aggregate data rates from large arrays remain within the transfer and processing budget of downstream analysis~\cite{Tahvildari2015, Zhang2015, Jain2017, Tahvildari2017, Rahman2021, Jones2025}. 
A practical consideration is that the RA and MM parameters require some prior knowledge of the expected analyte timescale; in practice, this constraint is mild, the parameter space is forgiving, and settings transfer robustly across conditions once established.
 However, excessive smoothing to suppress noise can attenuate the amplitude of fast translocations, reducing sensitivity for short-lived species. Furthermore, the system load is governed by the event capture rate, which increases memory-controller utilization. In high-concentration experiments, data reduction efficiency naturally diminishes as informative signals occupy the majority of the trace.

The biological implications will be fully realized as arrays scale to hundreds of individually addressable pores, enabling the molecule counts and statistical power that single-molecule diagnostics demand~\cite{Ratinho2025}. High-bandwidth parallel readout is particularly valuable where analyte information is sparse: free protein translocations under native conditions report on structural dynamics and conformational heterogeneity~\cite{Schmid2021}, denatured proteins interacting with the pore substrate yield sequence fingerprints~\cite{soni2025full}, and DNA-based molecular carriers or structural barcodes encode target binding events for RNA or protein detection~\cite{Xue2020, Bell2016, bovskovic2023simultaneous}. 
Designer nanoparticles, such as the NANPs demonstrated here, extend the approach toward therapeutic screening and biomarker quantification~\cite{alibakhshi2017picomolar, pandey2025chemical}. 
Transitioning to large-scale applications requires scaling the hardware: integrating  EUV photolithographic nanopore fabrication~\cite{Chaudhuri2025} with individually addressable fluidics will provide the dense arrays needed for population-level analysis. Integrating machine-learning classification directly into the acquisition hierarchy will then push the edge-computing paradigm toward adaptive, physics-informed triggering and real-time molecular identification at scale.

\newpage

\section{Methods}

Data Sieving is a modular acquisition architecture that integrates real-time filtering, triggering, and selective data retention directly into the measurement pipeline, replacing the conventional paradigm of indiscriminate recording followed by offline analysis. 
Rather than writing all raw data to disk, Data Sieving identifies event candidates in real time and preserves only short, full-resolution snapshots surrounding molecular translocation events. Raw snapshots are stored without modification, ensuring compatibility with any downstream analysis pipelines. The architecture consists of four sequential building blocks: high-bandwidth acquisition, event-candidate detection, active feedback, and event pruning. Each is implemented for parallel execution across multiple nanopore channels, enabling real-time operation at MHz sampling rates without computational bottlenecks or limitations on channel count, as illustrated in Fig.~\ref{fig:fig1}a.

\subsection{Nanopores and Electronics}

All solid-state nanopores were purchased from Norcada with a nominal diameter of 10\,nm. Ionic current measurements were performed using integrated electronic amplifiers from Elements~Srl., including their single-channel, 4-channel, 16-channel, and 24-channel systems. More information about the nanopores and the electronics is available in Supplementary Material Section~\ref{sec:sm_poredata}.

\subsection{Instrumentation and Software Architecture}

Data Sieving is governed by a custom-engineered software stack for acquisition and control, developed in NI LabVIEW and NI TestStand. To ensure scalability across high-bandwidth multi-channel arrays, the instrumentation is organized into a three-tier decoupled architecture (Fig.~\ref{fig:fig1}) that separates the user interface and test logic from the high-frequency data processing kernels.

The software stack is divided into three functional abstraction layers:
\begin{itemize}
    \item \textbf{Layer 1: User Interface (UI):} A LabVIEW-based interface provides real-time visualization of ionic current traces and experimental metadata, enabling immediate operator feedback and manual process intervention.
    \item \textbf{Layer 2: Test Executive:} Automated experimental workflows and sequencing are managed by NI TestStand. This executive layer standardizes test management, error handling, and automated report generation, ensuring reproducible experimental conditions across different nanopore chips. 
    \item \textbf{Layer 3: High-Throughput Processing:} This layer manages direct hardware communication and high-speed data orchestration. By utilizing highly optimized compiled code, this layer minimizes CPU-side overhead, ensuring the system can handle the data deluge inherent to MHz-rate sampling. 
\end{itemize}

To maintain real-time performance, the instrumentation stack minimizes CPU data handling. Upon arrival from device drivers, raw ionic current data is streamed directly to the GPU for processing. All signal processing algorithms, including the rolling-average (RA) filter and windowed min-max (MM) detection, are implemented using the G2CPU High-Performance Computing (HPC) Toolkit for LabVIEW.

Only identified event candidates and downsampled baseline metadata are transferred back to the CPU host.  This data is then used for:
\begin{enumerate}
    \item \textbf{Structured Storage:} Raw binary snapshots are archived in the Technical Data Management Streaming (TDMS) format.
    \item \textbf{Active Feedback:} The system monitors pore health and issues real-time commands for automated declogging. 
\end{enumerate}

This modular architecture guarantees that hardware upgrades or algorithmic refinements can be integrated without a fundamental redesign of the core acquisition pipeline, maintaining scalability as channel counts increase. Additional information is available in Supplementary Material Section~\ref{sec:sm_instrumentation}.

\subsection{Data Storage}

Data is stored using the Technical Data Management Streaming (TDMS) format. TDMS format provides a hierarchical structure (file, groups, channels) that supports efficient storage of raw event snapshots, downsampled baseline traces, and channel-specific metadata. A detailed description of the TDMS layout used in this work is provided in Supplementary Material Section~\ref{sm:file_type}.

\subsection{Event Candidate Detection}

The architecture of the Data Sieving framework is illustrated in Fig.\ \ref{fig:fig1}a. Measurements were conducted using multichannel platforms from Elements Srl (1, 4, 16, or 24 channels) featuring integrated transimpedance amplifiers (TIAs) and analog-to-digital converters (ADCs). 
Raw high-frequency data is streamed via a dedicated application programming interface to a custom LabVIEW-based instrumentation package, where snapshots from each channel are processed using a numerically optimized triggering algorithm that offloads the computational burden to the GPU. This edge-computing approach allows the system to reject the vast majority of non-informative baseline data in real time, passing only flagged event candidates for further processing and permanent storage.
The parallelized detection sequence is shown in Fig.\ \ref{fig:fig1}b. Translocation events are identified while data is streamed into a circular buffer for lossless recovery. The system employs a two-stage trigger consisting of a rolling-average (RA) filter followed by a windowed min–max (MM) detector. The RA stage suppresses high-frequency noise to produce a smoothed trace suitable for rapid evaluation, while the MM stage computes the peak-to-peak amplitude difference (max minus min) within fixed windows, as shown in Fig.\ \ref{fig:fig1}c. Windows exceeding a channel-specific threshold $T$ are flagged as potential events, triggering the retrieval and storage of the corresponding high-resolution raw segments.
These thresholds are automatically initialized from baseline noise statistics at the onset of acquisition but can be manually overridden to accommodate specific experimental conditions. 
This RA+MM scheme is computationally lightweight and embarrassingly parallel; each window and channel can be processed independently, enabling deployment across large nanopore arrays without saturating compute resources. 
Furthermore, because detection decisions are based on local peak-to-peak variations rather than absolute current levels, the system remains robust to slow baseline drift. Detailed mathematical derivations of the algorithm and parameter selections are provided in Supplementary Material Section~\ref{sm:event_candidate}.

\subsection{Active Feedback and Fast Actuation}

Beyond event detection, Data Sieving maintains a downsampled representation of the pore baseline that serves as a lightweight health monitor for each channel. Moreover, the RA–MM detection stream provides a real-time estimate of the pore state, allowing the system to implement closed-loop actuation protocols. 
Sustained deviations from the open-pore baseline are interpreted as potential clogging or instability, which triggers an automated polarity inversion on the affected channel to restore open-pore conductance.

This actuation is asynchronous and does not interrupt acquisition on adjacent channels. If the pore fails to recover after a pre-defined number of attempts, the channel is temporarily disabled to maintain overall system stability and prevent unnecessary I/O bus occupancy. Automatic declogging represents one instance of this active-feedback framework; the same architecture can support additional closed-loop operations, such as adaptive bias adjustment, dynamic threshold tuning, or channel-specific duty cycling. Further details regarding the active-feedback logic are provided in Supplementary Material Section~\ref{sm:declog}.

\subsection{Event Pruning}

To further reduce data volume and the computational load of downstream analysis, raw event candidates undergo a second stage, which we call 'event-pruning', as illustrated in Fig.\ \ref{fig:fig1}d. Candidate snapshots initially include baseline padding to ensure complete capture. The pruning sequence begins with a digital low-pass filter and a voltage-stability check to identify non-stationary segments or hardware artifacts.

For stable segments, event onset and offset boundaries are localized using two independent, parallel algorithms: (1) a cumulative sum (CUMSUM) change-point detector and (2) a statistical metric extraction approach that identifies the shortest interval where the signal deviates from the baseline. If both methods reach consensus, the event is trimmed to the agreed interval. If the methods diverge or the bias is unstable, the full, untrimmed snapshot is retained to prevent information loss. 
While pruning provides a marginal benefit in total data volume reduction, it significantly reduces the number of data points for downstream real-time processing, thereby lowering the system's total computational load. More technical details and parameter selections for this stage are provided in Supplementary Material Section~\ref{sm:pruning}.

\newpage

\section*{Authorship contribution statement}

M.C.\ and N.B.\ contributed equally to this work.  
M.C.\, N.B.\, W.P.\, E.B.\, and S.M.\ conceived and designed the experiments.
N.B.\ developed the instrumentation.  
M.C.\ performed the data analysis. 
M.C.\ and W.B.\ developed general data processing tools. 
W.P.\, E.B.\, N.B.\, W.R.\, L.V.\, and M.C.\ conducted the experiments.
E.S.\, and K.A.A.\ provided the DNA/RNA constructs.
All authors except E.S.\, and K.A.A.\, contributed to shaping the concept and developing the system.  
S.M.\ provided overall project leadership and supervision.  
M.C.\ and S.M.\  drafted the manuscript with input from all authors.  
All authors reviewed and approved the final version of the manuscript.

\section*{Acknowledgments}
The authors thank Federico Thei and Elements Srl. The authors also thank Silvia Lenci, Ashesh Chaudhuri, Ananth Subramanian, and Charlotte D'Hulst.
The NANP related research reported in this publication was supported in part by the National Institute of General Medical Sciences of the National Institutes of Health under Award Number R35 GM139587 (to K.A.A.). The content of this publication does not necessarily reflect the views or policies of the Department of Health and Human Services, nor does mention of trade names, commercial products, or organizations imply endorsement by the US Government.

\section*{Declaration of competing interest}
The authors declare a competing interest: a patent application covering aspects of the work reported in this manuscript has been submitted. No other competing interests are declared.

\bibliography{biblio}

\begin{tocentry}
    \centering
\includegraphics[width=7cm]{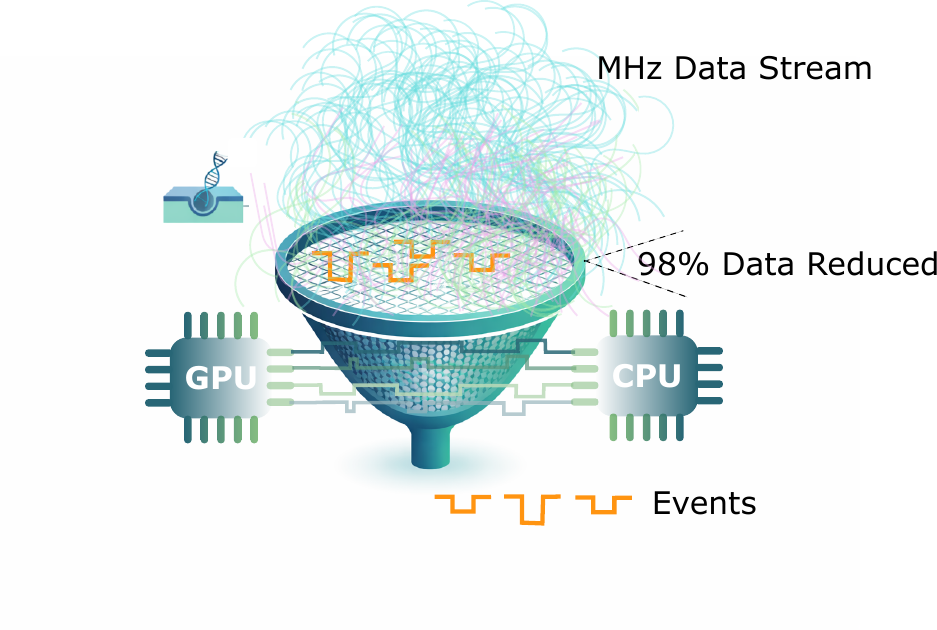}
\end{tocentry}

\end{document}